\documentclass[pdflatex,sn-mathphys-num]{sn-jnl}

\usepackage{graphicx}%
\usepackage{multirow}%
\usepackage{amsmath,amssymb,amsfonts}%
\usepackage{amsthm}%
\usepackage{mathrsfs}%
\usepackage[title]{appendix}%
\usepackage{xcolor}%
\usepackage{textcomp}%
\usepackage{manyfoot}%
\usepackage{booktabs}%
\usepackage{siunitx}%

\usepackage{algorithm}%

\usepackage{algorithmicx}%
\usepackage{algpseudocode}%
\usepackage{listings}%

\theoremstyle{thmstyleone}%
\theoremstyle{thmstyletwo}%

\theoremstyle{thmstylethree}%

\begin{document}

\title[Article Title]{Jet flavor tagging with Particle Transformer for Higgs factories }


\author*[1,2]{\fnm{Taikan} \sur{Suehara}}\email{suehara@icepp.s.u-tokyo.ac.jp}

\author[2]{\fnm{Takahiro} \sur{Kawahara}}

\author[3]{\fnm{Tomohiko} \sur{Tanabe}}

\author[2]{\fnm{Risako} \sur{Tagami}}

\affil*[1]{\orgdiv{International Center for Elementary Particle Physics}, \orgname{The University of Tokyo}, \orgaddress{\street{7-3-1 Hongo, Bunkyo-ku}, \city{Tokyo}, \postcode{133-0033}, \country{Japan}}}

\affil[2]{\orgdiv{Department of Physics, Graduate School of Science}, \orgname{The University of Tokyo}, \orgaddress{\street{7-3-1 Hongo, Bunkyo-ku}, \city{Tokyo}, \postcode{133-0033}, \country{Japan}}}

\affil[3]{\orgname{MI-6 Ltd.}, \orgaddress{\street{8-13 Nihombashi-Kobunacho, Chuo-ku}, \city{Tokyo}, \postcode{103-0024}, \country{Japan}}}


\abstract{We study the performance of the Particle Transformer (ParT) for jet flavor tagging using ILD full simulation events (1M jets) as well as fast simulation samples (10M and 1M jets). We perform 3-category ($b/c/d$), 6-category ($b/c/d/u/s/g$), and 11-category trainings (including quark--antiquark separation), incorporating multivariate hadron particle identification information from $dE/dx$ and time-of-flight. For $b$/$c$ tagging, we observe a factor of 5--10 times better rejection of different flavors at the same efficiency over previous BDT-based taggers, and we obtain reasonable performance for strange tagging and quark/antiquark separation. The 10M-jet fast simulation study indicates that further gains are possible with higher training statistics.}




\maketitle

\begin{center}
{\small Proceedings No.~ILD-PHYS-PROC-2026-010

Talk presented at the International Workshop on Future Linear Colliders (LCWS 2025), October 20-24 2025. This work was carried out in the framework of the ILD Concept Group.
}
\end{center}

\section{Introduction}\label{sec1}

Event reconstruction at future $e^+e^-$ colliders relies on extracting fine-grained information from highly segmented tracking and calorimeter systems and translating it into physics objects such as charged particles, photons, and jets. As detector granularity increases, pattern recognition challenges become more prominent, making modern machine learning (ML) approaches a natural choice for several reconstruction and identification tasks.

Jet flavor tagging aims to identify the initiating parton flavor (e.g., $b$, $c$, or light flavor) using information carried by reconstructed particles inside a jet. A key handle for heavy flavor tagging is the finite lifetime of heavy flavor hadrons, which leads to displaced tracks and secondary vertices. Traditional approaches for Higgs factory studies, such as LCFIPlus\cite{LCFIPlusPaper}, explicitly reconstruct secondary vertices and then combine vertex- and track-level observables in multivariate classifiers.

More recently, end-to-end deep learning models have been developed that ingest low-level per-particle (or per-track) information and learn powerful representations without relying on a dedicated vertex reconstruction step. For example, ParticleNet, which operates on point-cloud representations and uses graph-based architectures, has been applied to fast simulation samples for FCC-ee and was reported to achieve substantially improved $b$-tag rejection compared to LCFIPlus\cite{FCCPN}. Similar studies of flavor identification using ParticleNet-based algorithms have also been reported in the context of CEPC, another proposed $e^+e^-$ Higgs factory\cite{joi_cepc}.
Transformer-based models have also demonstrated strong performance in LHC jet tagging benchmarks; in particular, the Particle Transformer (ParT)\cite{Qu:2022mxj} leverages attention mechanisms that scale well with large training samples.

In this work, we study the performance of ParT for jet flavor tagging using ILD full simulation events as well as large fast simulation samples. We focus on Higgs factory conditions and assess the impact of detailed detector information and particle identification (PID) inputs, which are expected to be especially relevant for strange jet tagging.

\section{Simulation and Data samples}\label{sec2}

We use simulated events for the International Large Detector (ILD) concept\cite{ILD:2020qve}. The ILD tracking system comprises a silicon vertex detector, silicon inner/outer trackers, and a central Time Projection Chamber (TPC) in the barrel region, complemented by silicon based tracking in the forward region. The calorimeter system is highly granular, enabling Particle Flow reconstruction, and is surrounded by a 3.5~T solenoid; muon detectors are located outside the magnet.

Several detector features provide PID capability. In the TPC, specific energy loss ($\mathrm{d}E/\mathrm{d}x$) measurements allow charged hadron separation over a broad momentum range. In addition, we assume timing information in the first ten layers of the electromagnetic calorimeter and use time-of-flight information for PID, with an indicative single hit time resolution of 100~ps.

The event processing is based on the iLCSoft\cite{iLCSoft_github} stack, with Geant4-based detector simulation using the ILD detector description (DDSim), digitization that models detector effects, and low-level tracking reconstruction. For the jet-based analyses, we reconstruct Particle Flow objects with PandoraPFA\cite{Thomson:2009rp} and form jets using the Durham clustering algorithm. The reconstructed particles associated with each jet are used to build the model inputs.
As part of the reconstruction chain, we apply the Comprehensive PID algorithm\cite{cpid}, which combines $\mathrm{d}E/\mathrm{d}x$ and time-of-flight related observables in a boosted decision tree to classify charged particles. The PID performance on $ZH \to \nu\nu ss$ sample is shown in Table \ref{tab:pid_confusion_pgt5}. The resulting PID responses are used as additional per-particle features for the jet taggers and are expected to provide extra discriminating power, particularly for strange jet identification.

\begin{table}[htbp]
\centering
\caption{Confusion matrix for tracks with $p > 5$~GeV with CPID, evaluated by $ZH \to \nu\nu ss$ events. Official weights of 2fZhad\_12bins\_conservative\cite{CPID_ILDConfig} are used for the CPID.}
\label{tab:pid_confusion_pgt5}
\begin{tabular}{lccccc}
\hline
 & true $K$ & true $\pi$ & true $p$ & true $e$ & true $\mu$ \\
\hline
$K$   & 0.74 & 0.07 & 0.20 & 0.13 & 0.16 \\
$\pi$  & 0.07 & 0.89 & 0.03 & 0.40 & 0.37 \\
$p$    & 0.18 & 0.03 & 0.76 & 0.09 & 0.06 \\
$e$    & 0.00 & 0.00 & 0.00 & 0.38 & 0.01 \\
$\mu$  & 0.01 & 0.01 & 0.00 & 0.01 & 0.40 \\
\hline
\end{tabular}
\end{table}

Our baseline physics process is $e^+e^- \to ZH \to \nu\nu\, q\bar{q}$ at $\sqrt{s}=250$~GeV. To evaluate flavor tagging performance across Higgs decays, we consider six Higgs-to-parton categories, $H\to b\bar{b},\, c\bar{c},\, s\bar{s},\, u\bar{u},\, d\bar{d},\, gg$;\footnote{Although the branching fractions for Higgs decays into light quarks are very small, in this sample the decay modes are forced so that these events can be used as flavor tagging samples.}in the full simulation samples, approximately $2\times10^6$ jets are available per category.
We use the sample prepared by ILD central production (mc-2020\cite{mc2020}) with iLCSoft v02-02. Event generation was done by Whizard 2.8.5\cite{whizard}, hadronized by Pythia 6.427\cite{pythia6}.
To study scaling behavior and enable analyses requiring much larger statistics, we also use samples produced privately with Whizard 3.1.5 hadronized by Pythia 6 with the same configuration as the full simualtion and SGV\cite{sgv} fast simulation, comprising roughly $1\times10^7$ jets per category for $H\to b\bar{b},\, c\bar{c},\, d\bar{d}$ categories. The SGV samples are only used for heavy flavor ($b/c$) discrimination since the SGV configuration employed here does not include hadron-level PID information that is essential for strange tagging.


\section{Algorithm}

We employ the Particle Transformer (ParT) as the baseline jet flavor tagging model. ParT follows the standard Transformer design with an embedding stage, a stack of self-attention blocks, and a final multilayer perceptron for classification. In addition to the usual content-based attention, ParT incorporates a pairwise ``interaction'' term derived from the kinematics of particle pairs, which is injected as a bias into the attention weights. This interaction-aware attention is particularly suitable for jet tagging, where correlations among constituents carry essential information.

Each jet is represented as a set of reconstructed particles and is processed as a variable length sequence. Charged and neutral particles are embedded with separate input projections to account for the different feature availability. For charged particles, the input features include impact parameter related variables, track parameter uncertainties, PID-related quantities, and kinematic variables; for neutral particles, only the available kinematic and PID-related features are used. The per-particle input variables used in this study are summarized in Table~\ref{tab:input_features}.

We train three classification settings with different label granularities. In the 3-category training, jets are labeled as $b$, $c$, or light (denoted as $d$) using samples generated for $H\to b\bar{b}$, $H\to c\bar{c}$, and $H\to d\bar{d}$. In the 6-category training, the labels are extended to $b$, $c$, $s$, $u$, $d$, and $g$ using samples generated for $H\to b\bar{b}$, $c\bar{c}$, $s\bar{s}$, $u\bar{u}$, $d\bar{d}$, and $gg$. For the 3- and 6-category trainings, we assign labels at the event level based on the generated sample type, without accessing per-event MC truth matching in the reconstruction.


In the 11-category training which we introduced anti-quark categories ($\bar{b}, \bar{c}, \bar{s}, \bar{u}, \bar{d}$ in addition to the 6 categories), we assign labels by matching reconstructed jets to the underlying partons using MC truth information and distinguish quarks from antiquarks. For each reconstructed particle, we first identify the corresponding MC truth particle and group truth particles by their color-singlet system without using angular information. We then resolve the parton assignment only within each color singlet by comparing the opening angle between the truth particle and the partons (quark/gluon) produced directly from the color singlet, and assign the truth particle to the closest parton. Using this particle--parton association, we label each reconstructed jet by the parton that receives the largest fraction of the reconstructed constituent energy.

For all trainings, we split each dataset into 80\% for training, 10\% for validation, and 10\% for testing.

\begin{table}[htbp]
\centering
\caption{Per-particle input variables for the Particle Transformer. (C) in Group indicates that the variables are only used for charged particles. Interaction (bias) variables are used for biasing attention weights and all others are for features input.}
\label{tab:input_features}
\small
\setlength{\tabcolsep}{4pt}

\begin{tabular}{p{0.2\linewidth}p{0.28\linewidth}p{0.45\linewidth}}
\hline
Group & Variables\cite{lcfiplus_release} & Description \\
\hline
Impact parameters (C) & $d_{xy},\, d_{z}$; sip2D, sip3D & Signed transverse/longitudinal impact parameters and their derived 2D/3D values and significances\\
Jet distances (C) & $d_j$ & Track displacement relative to the jet direction with value and significance\\
Particle ID & Muon/Electron/Gamma flags; charged/neutral hadron flags; type & PID indicator variables and reconstructed PDG ID\\
Particle ID (C) & CPID outputs ($\pi$, $K$, $p$ probabilities) & Comprehensive PID (BDT) outputs (used in 6- and 11-category trainings)\\
Kinematics & Charge; $\log(E/E_{\mathrm{jet}})$; $\Delta\theta$; $\Delta\phi$ & Per-particle kinematic variables relative to the jet axis\\
Track errors (C) & $\sigma$ (covariance elements) & Track parameter uncertainty information\\
\hline
Interaction (bias) & $\log k_t$, $\log z$, $\log \Delta R^2$, $\log m^2$ & Pairwise variables used to construct the interaction bias in attention\\
\hline
\end{tabular}

\end{table}





\section{Results}\label{sec:results}

\begin{figure}[t]
\centering
\begin{minipage}[t]{0.45\textwidth}
\centering
\includegraphics[width=\linewidth]{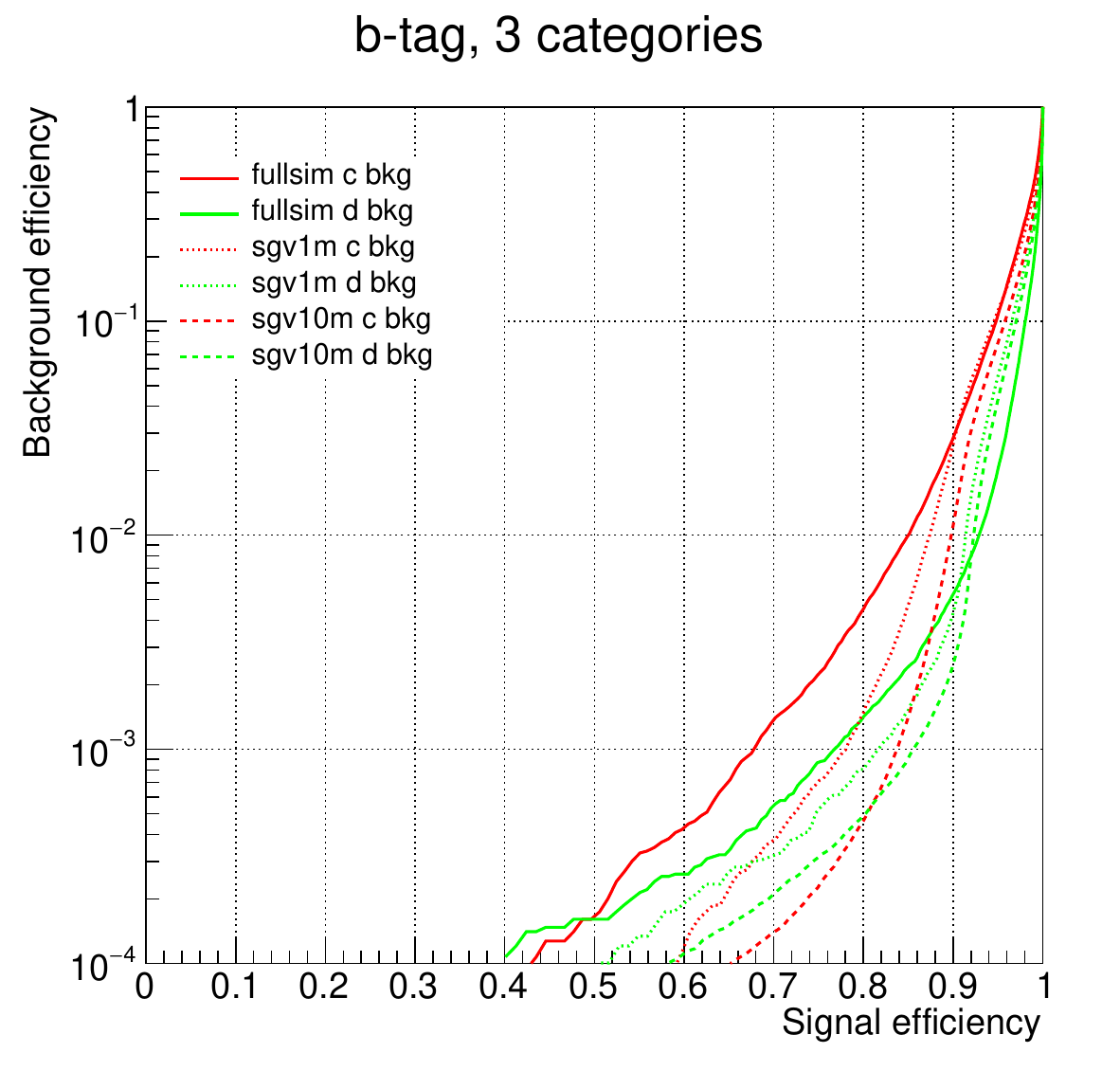}
(a) 3-cat.~$b$-tag
\end{minipage}\hfill
\begin{minipage}[t]{0.45\textwidth}
\centering
\includegraphics[width=\linewidth]{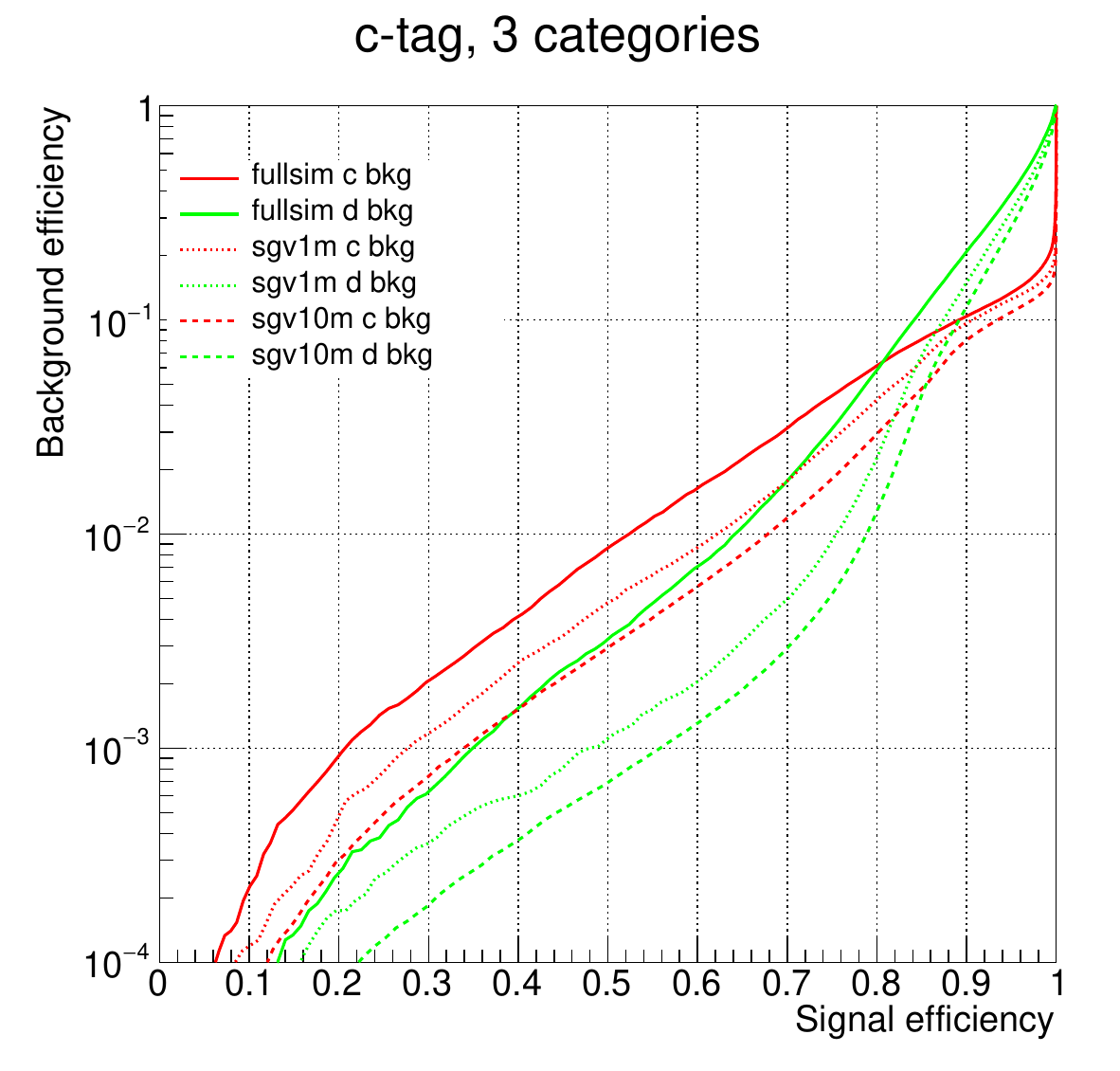}
(b) 3-cat.~$c$-tag
\end{minipage}

\begin{minipage}[t]{0.6\textwidth}
\centering
\includegraphics[width=\linewidth]{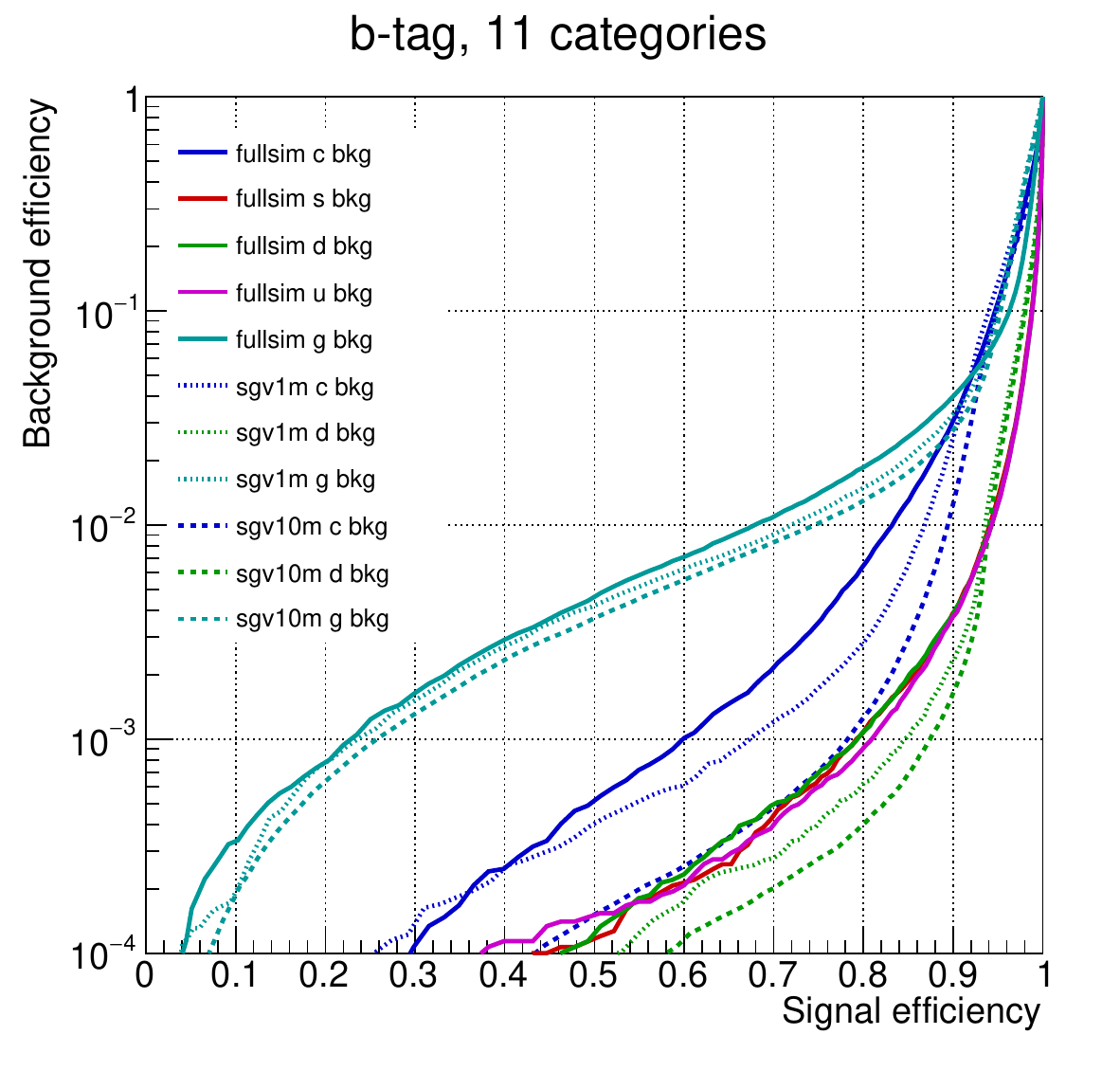}
(c) 11-cat.~$b$-tag
\end{minipage}

\caption{Performance plots for $b$- and $c$-tagging. The line colors indicate background jet categories. The solid, dotted, and dashed curves correspond to ILD full simulation, SGV fast simulation trained with 1M jets, and SGV fast simulation trained with 10M jets, respectively. (c) shows $b$-tagging of 11 categories where $s$ and $u$ background of SGV are omitted because they are essentially same as $d$ without PID variables.}
\label{fig:btag_ctag}
\end{figure}

Figure~\ref{fig:btag_ctag} summarizes the $b$- and $c$-tagging performance for the 3-category and 11-category trainings. Compared to the conventional LCFIPlus approach, we observe a substantial reduction in background acceptance at a fixed signal efficiency. In particular, at a $b$-tag efficiency of 80\%, the acceptance for $c$-jet background is reduced to the $\mathcal{O}(10^{-3})$ level, and the light-flavor background to the $\mathcal{O}(10^{-3})$--$\mathcal{O}(10^{-4})$ level, depending on the training setup and simulation model, to be compared with LCFIPlus result of $6.3 \times 10^{-2}$ for $c$-jet background and $7.9 \times 10^{-3}$ for light-flavor background with the same 80\% efficiency working point.

The comparison between full simulation and SGV fast simulation indicates non-negligible differences in absolute performance, which we are currently investigating in more detail. In addition, within SGV we observe an improvement when increasing the training statistics from 1M to 10M jets. The 3-category and 11-category trainings yield broadly comparable $b$-tagging performance; however, in the high-purity regime the rejection of gluon jets becomes challenging in the 11-category setup, which is expected in part due to $g\to b\bar{b}$ splittings that can mimic heavy flavor signatures. For $c$-tagging, the task remains more difficult than $b$-tagging, yet the background acceptance at a $c$-tag efficiency of 50\% is still significantly improved compared to LCFIPlus.
Table~\ref{tab:eff_08}(a) lists representative working points extracted from Fig.~\ref{fig:btag_ctag} at a signal efficiency of 0.8, facilitating a direct comparison across full simulation and SGV with different training statistics. The SGV sample trained with 1M jets shows slightly better performance than the full simulation, which suggests that SGV does not perfectly reproduce the full-simulation response. The SGV result with 10M training jets is significantly better than that with 1M jets, indicating that training statistics are important for achieving high performance. Comparing the 3-category and 11-category trainings, the 3-category setup gives better results for $b$ tagging against $c$-jet background, suggesting that the 3-category weights are more directly optimized for this specific task. By contrast, for $b$ tagging against $d$-jet background, the 11-category setup performs better. This may be related to the fact that $u$ jets, which share similar properties with $d$ jets, are included in the training sample, effectively increasing the available statistics. In multiclass classification, it is therefore important to keep in mind that the relative statistics of the training samples can affect the performance.

\begin{table}[t]
\centering
\caption{(a) Background efficiencies at signal efficiency $=0.8$. Category shows 3-c (3-category) or 11-c (11-category) training, followed by signal -- background (eg.~$b$--$c$ means $b$-tag with $c$-background).
(b) Background efficiencies at signal efficiency $=0.8$ on strange tag with 6-category training and ILD full-simulation.}
\label{tab:eff_08}
\small\setlength{\tabcolsep}{4pt}

\begin{tabular}{cc}
\begin{minipage}[t]{0.62\linewidth}
\centering
(a)\\
\vspace{0.5em}
\begin{tabular}{lrrr}
\toprule

Category & Full sim & SGV (1M) & SGV (10M)\\
\midrule
3-c $b$ -- $c$ & $4.53\times10^{-3}$ & $1.48\times10^{-3}$ & $4.64\times10^{-4}$\\
3-c $b$ -- $d$ & $1.42\times10^{-3}$ & $8.05\times10^{-4}$ & $4.93\times10^{-4}$\\
3-c $c$ -- $b$ & $6.12\times10^{-2}$ & $4.24\times10^{-2}$ & $2.94\times10^{-2}$\\
11-c $b$ -- $c$ & $6.47\times10^{-3}$ & $2.83\times10^{-3}$ & $1.26\times10^{-3}$\\
11-c $b$ -- $s$ & $1.08\times10^{-3}$ & $5.81\times10^{-4}$ & $3.99\times10^{-4}$\\
11-c $b$ -- $d$ & $1.08\times10^{-3}$ & $6.20\times10^{-4}$ & $4.01\times10^{-4}$\\
11-c $b$ -- $u$ & $9.09\times10^{-4}$ & $5.99\times10^{-4}$ & $4.22\times10^{-4}$\\
11-c $b$ -- $g$ & $1.86\times10^{-2}$ & $1.49\times10^{-2}$ & $1.30\times10^{-2}$\\
\bottomrule
\end{tabular}
\end{minipage}\hfill
\begin{minipage}[t]{0.35\linewidth}
\centering
(b)\\
\vspace{0.5em}
\begin{tabular}{lS[table-format=2.2,minimum-decimal-digits=0]}
\toprule

\multicolumn{2}{c}{Efficiency ($\times 10^{-2}$)}\\
\midrule
$b$ bkg & 1.14\\
$c$ bkg & 12.9\\
$u$ bkg & 42.5\\
$d$ bkg & 40.5\\
$g$ bkg & 25.5\\
\bottomrule
\end{tabular}
\end{minipage}

\end{tabular}
\end{table}

\begin{figure}[t]
\centering
\begin{minipage}[t]{0.4\textwidth}
\centering
\includegraphics[width=\linewidth]{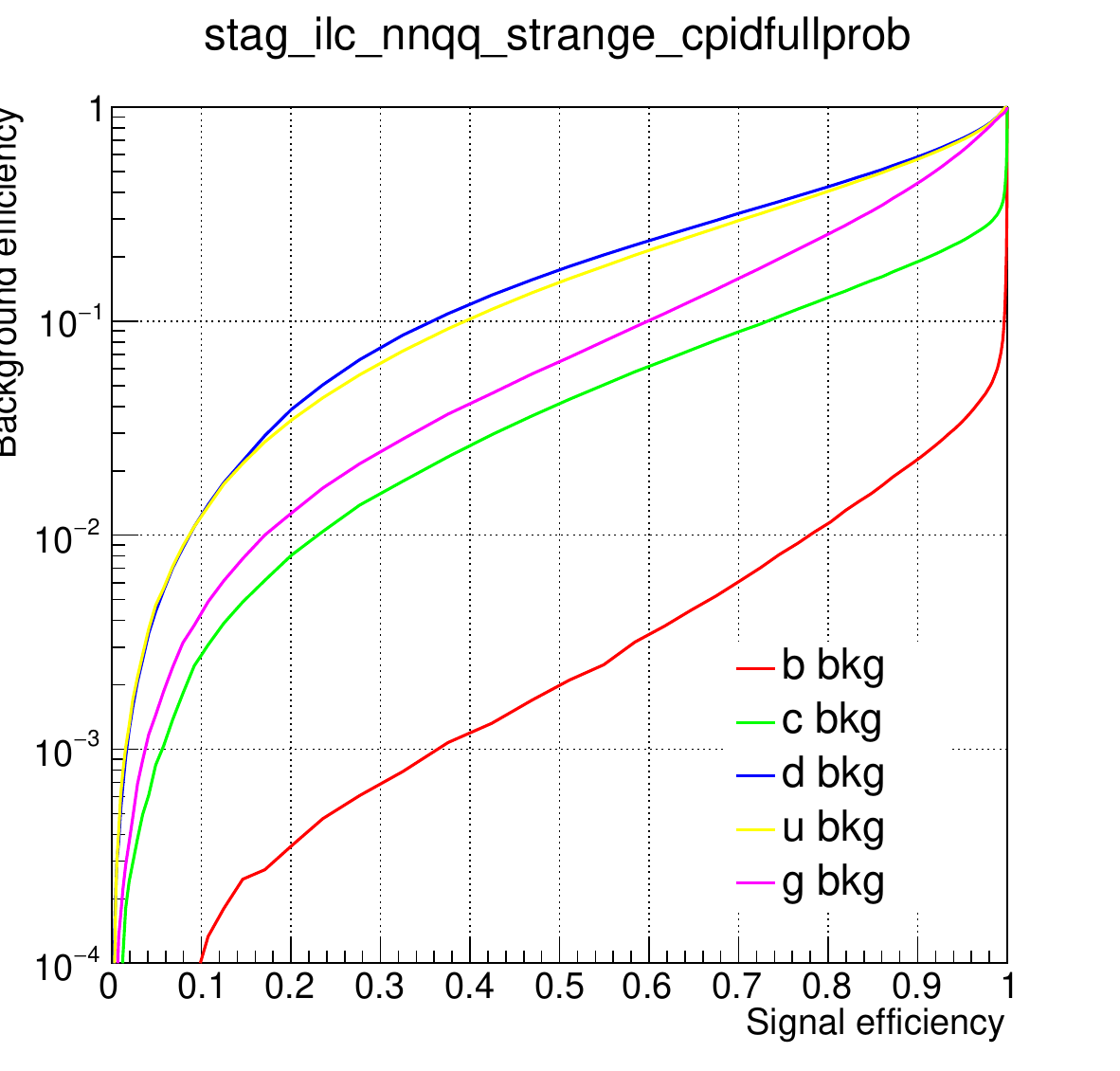}
(a)
\end{minipage}\hfill
\begin{minipage}[t]{0.55\textwidth}
\centering
\includegraphics[width=\linewidth]{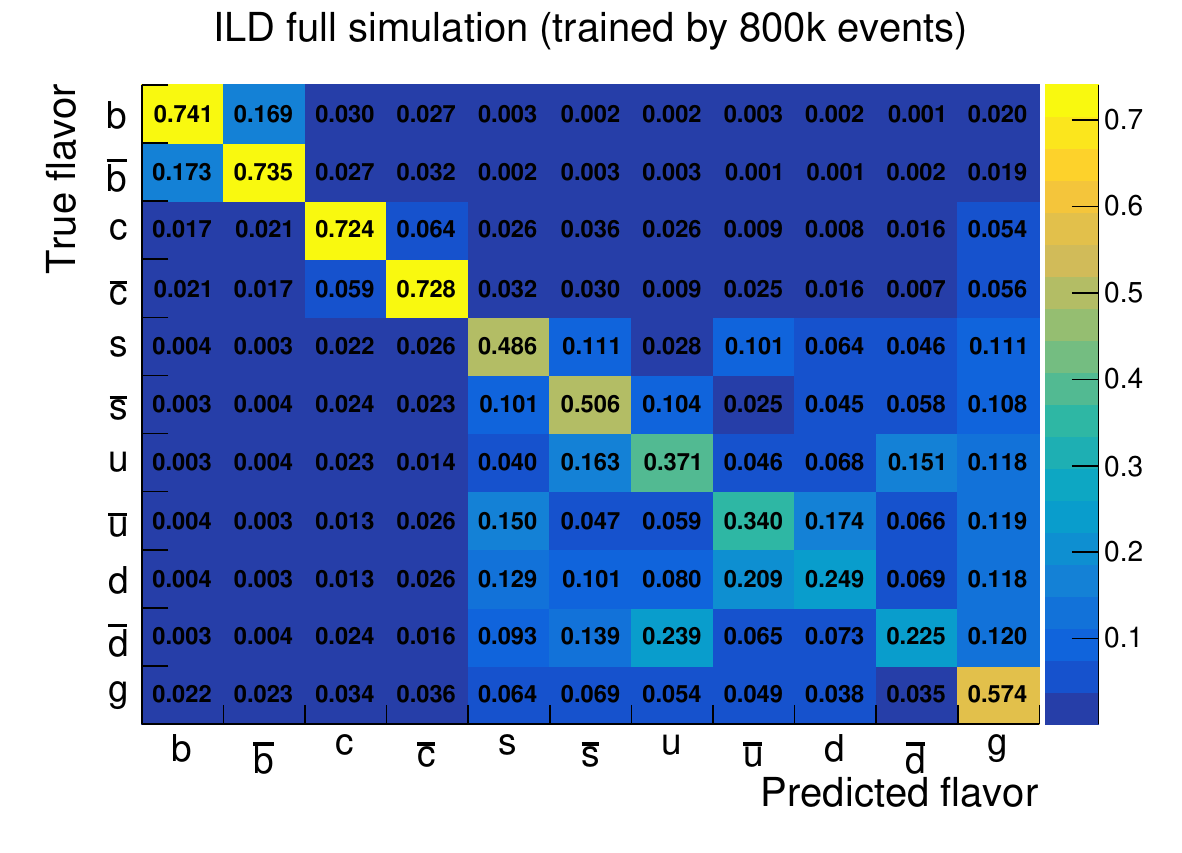}
(b)
\end{minipage}
\caption{(a) Signal-versus-background performance for 6-category strange tagging in ILD full simulation. (b) Confusion matrix for the 11-category classifier in ILD full simulation, where the predicted flavor is defined as the class with the highest predicted probability.}
\label{fig:stag_matrix11}
\end{figure}

Figure~\ref{fig:stag_matrix11}(a) shows the strange tagging performance in the 6-category training with full simulation. The strange background efficiencies at the working point of signal efficiency $\varepsilon_s=0.8$ (extracted from Fig.~\ref{fig:stag_matrix11}(a)) are summarized in Table~\ref{tab:eff_08}(b). Strange tagging relies strongly on hadron-level PID information inside the jet; however, jets initiated by non-strange partons may still contain strange hadrons through fragmentation, including contributions from gluon splitting. Consequently, the separation is intrinsically probabilistic, and we find that discriminating among $u$, $d$, and $g$ jets is particularly challenging. Gluon jets can nevertheless be separated from quarks to some extent due to their different kinematic and radiation patterns.


Figure~\ref{fig:stag_matrix11}(b) presents the agreement between the predicted labels (taken as the highest probability class) and the truth labels in the 11-category training. Beyond jet flavor identification, we observe that the model can separate quarks from antiquarks for heavy flavors ($b$, $c$, and $s$) with meaningful accuracy. The separation is strongest for charm, where the larger electric charge of the underlying hadrons provides additional information. In contrast, separating $u$ and $d$ (and their antiparticles) remains difficult, and in particular the discrimination of charge-conjugate light flavor pairs (e.g., $u$ vs.\ $\bar{d}$ and $d$ vs.\ $\bar{u}$) is essentially consistent with random guessing.

\section{Summary and Prospects}

We evaluated jet flavor tagging performance with the Particle Transformer using both ILD full simulation and large statistics fast simulation samples. For heavy flavor discrimination, across the 3-, 6-, and 11-category trainings, the ParT-based taggers consistently outperform the conventional baseline, achieving substantially lower background acceptance at fixed signal efficiency (Fig.~\ref{fig:btag_ctag} and Table~\ref{tab:eff_08}(a)). For strange tagging, we find that incorporating hadron-level PID information (CPID) provides useful discrimination, although the separation among light flavors remains intrinsically limited by fragmentation (Fig.~\ref{fig:stag_matrix11}(a) and Table~\ref{tab:eff_08}(b)). In the 11-category setting, the model further demonstrates sensitivity to quark--antiquark separation for heavy flavors, as reflected in the truth--prediction agreement matrix (Fig.~\ref{fig:stag_matrix11}(b)). These results show similarities to those reported for FCC-ee\cite{FCCPN} and CEPC\cite{joi_cepc}, although a detailed comparison is not practical because of the differences in the conditions of the simulation and analysis setups.

The SGV studies indicate a clear dependence on training statistics, motivating further investigations with even larger samples and a systematic validation with full simulation. The resulting taggers can be deployed in the ILD analysis workflow through an ONNX-based inference pipeline\cite{onnxruntime}, and updates of physics analyses using this tool are ongoing\cite{lcws2025-seino}\cite{lcws2025-bryan}. Future work will focus on improving the input representation and training strategy, as well as quantifying systematic differences between full and fast simulation and their impact on physics measurements.

\section*{Acknowledgements}
We would like to thank the LCC generator working group and
the ILD software working group for providing the simulation and reconstruction tools and
producing the Monte Carlo samples used in this study. This work has benefited
from computing services provided by the ILC Virtual Organization, supported by the national
resource providers of the EGI Federation and the Open Science GRID.
This work is supported by JSPS KAKENHI Grant Number JP22H05113.


\bibliography{sn-bibliography}

\end{document}